\documentclass[twocolumn]{aastex61}

\received{XXXX}
\revised{XXXX}
\accepted{XXXX}

\usepackage{natbib}
\usepackage{comment}
\usepackage{amsmath}
\usepackage{float}

\begin{document}

\title{\textit{K2} Ultracool Dwarfs Survey. IV. Monster flares observed on the young brown dwarf CFHT-BD-Tau 4.}

\correspondingauthor{Rishi R. Paudel}
\email{rpaudel@udel.edu}

\author{Rishi R. Paudel, John E. Gizis, D. J. Mullan}
\affiliation{Department of Physics and Astronomy, University of Delaware, Newark, DE, 19716, USA}
\author{Sarah J. Schmidt}
\affiliation{Leibniz-Institute for Astrophysics Potsdam (AIP), An der Sternwarte 16, 14482, Potsdam, Germany}
\author{Adam J. Burgasser}
\affiliation{Center for Astrophysics and Space Science, University of California San Diego, La Jolla, CA 92093, USA}
\author{Peter K. G. Williams}
\affiliation{Harvard-Smithsonian Center for Astrophysics, 60 Garden Street, Cambridge, MA 02138, USA}
\author{Edo Berger}
\affiliation{Harvard-Smithsonian Center for Astrophysics, 60 Garden Street, Cambridge, MA 02138, USA}

\begin{abstract}
We present photometric measurements of two superflares observed on a very young brown dwarf CFHT-BD-Tau 4, observed during Campaign 13  of the \textit{Kepler K2} mission. The stronger of the two superflares brightened by a factor of $\sim$48 relative to the quiescent photospheric level, with an increase in \textit{Kepler} magnitude $\Delta \tilde{K_{p}}$ = -4.20. It has an equivalent duration of $\sim$107 hour, a flare duration of 1.7 day, and an estimated total bolometric (ultraviolet/optical/infrared) energy up to 2.1 $\times$ 10$^{38}$ erg. The weaker of the two superflares is a complex (multipeaked) flare with an estimated total bolometric (UV/optical/IR) energy up to 4.7 $\times$ 10$^{36}$ erg. They are the strongest flares observed on any brown dwarf so far. The flare energies are strongly dependent on the value of visual extinction parameter $A_{V}$ used for extinction correction. If we apply a solar flare-model to interpret the two superflares, we find that the magnetic fields are required to be stronger by as much as an order of magnitude than previous reports of field measurements in CFHT-BD-Tau 4 by \cite{2009ApJ...697..373R}. On the other hand, if we interpret our data in terms of accretion, we find that the requisite rate of accretion for the stronger superflare exceeds the rates which have been reported for other young brown dwarfs.
\end{abstract}
\section{INTRODUCTION} \label{sec:introduction}
Rapidly rotating young solar mass stars in the Orion Nebula Cluster are capable of producing X-ray flares with energies in the range 10$^{34}$-10$^{36}$ erg \citep{2005ApJS..160..423W}. Similarly, low mass stars and young brown dwarfs (hereafter BDs) in Orion Nebula region and Taurus molecular cloud also have high X-ray emission. There is no significant difference in X-ray activity levels of the low mass stars and the young BDs with similar spectral types implying that X-ray activity levels are determined by effective temperatures rather than masses and surace gravities of (sub)stellar objects \citep{2005ApJS..160..582P,2007A&A...468..353G,2007A&A...468..391G,2008ApJ...688..418G,2008ApJ...688..437G}. Hence, young BDs can be magnetically active as the low mass stars, and produce huge flares. The young stars and BDs may have accreting or non-accreting disks which limit the X-ray flare loop sizes ($<$ Keplerian corotation radii) \citep{2008ApJ...688..437G}. In addition to the magnetic reconnection events which occur in diskless stars, the interaction of disks of young stars with their magnetospheres can also trigger large scale magnetic reconnection events which manifest as flares \citep{2009ApJ...694.1045Z}. \cite{2000A&A...356..949S} report the presence of an accretion disk makes no difference on X-ray flares on stars in clusters with ages 1-3 Myr. 
\\
\\
 The continuous time coverage of \textit{Kepler K2} mission \citep{2014PASP..126..398H} over months, has enabled us to study white light flare rates of several ultracool dwarfs (hereafter UCDs) which have spectral types $\gtrsim$ M6 and various ages \citep{2017ApJ...838...22G,2017ApJ...845...33G,2018arXiv180307708P}. Those flare rates will be helpful to understand the evolution of magnetic dyanmo in UCDs. The results of \cite{2017ApJ...845...33G} show that young brown dwarfs like 2MASS J03350208+2342356 (24 Myr old brown dwarf, hereafter 2M0335+2342)  and CFHT-PL-17 (a brown dwarf member of Pleiades) are capable of producing strong white light superflares with energies $>$ 10$^{33}$ erg. Superflares are thus ubiquitious in \textit{Kepler} G, K, M and L stars \citep{2012Natur.485..478M,2013ApJS..209....5S,2013ApJ...771..127N,2014ApJ...792...67C,2017ApJ...845...33G,2017ApJ...838...22G}.  The young brown dwarfs are in the process of contraction and have radii $>$0.5$R_{\odot}$ if they are few million years old. They are fully convective and have high luminosities. The energy flux scaling law predicts that strong magnetic fields are produced in young BDs and exoplanets \citep{2010A&A...522A..13R,2009Natur.457..167C}. This is supported by detection of 5 kG magnetic field on a 22 Myr, M8.5 brown dwarf LSR J1835+3259 \citep{2017arXiv170902861B}. \cite{2009ApJ...697..373R} however reported weak magnetic fields with strengths of few hundred gauss in four young ($\lesssim$10 Myr) accreting BDs with $v \sin i$ $>$ 5 km s$^{-1}$. The weaker fields may be due to presence of disk around such objects or that they do not follow scaling law \citep{2010A&A...522A..13R}. \\
\\ 
Here, we present the photometric measurements of two superflares observed on a very young brown dwarf CFHT-BD-Tau 4 (hereafter CT4). Discovered by \cite{2001ApJ...561L.195M}, the presence of disk around CT4 makes it more valuable for studies regarding planet formation around low mass stars and BDs \citep{2014ApJ...791...20R}. We present the photometric and physical properties of CT4 in Section \ref{sec:target properties}. In Section \ref{sec: data reduction and computations}, we present the data reduction, flare photometry and flare energy computation, and discuss the results in Section \ref{sec:discussion}. 
\section{Target Characteristics} \label{sec:target properties}
CFHT-BD-Tau 4 (2MASS J04394748+2601407) is a young M7 brown dwarf in the Taurus star-forming region, with an estimated age $\sim$1 Myr old \citep{2017AJ....153...46L} and at a distance of 147.1$\pm$5.2 pc \citep{2016A&A...595A...1G,2018arXiv180409365G}. It has an effective temperature (T$_{eff}$) equal to 2900 K and bolometric luminosity ($L_{bol}$) equal to 0.03 $L_{\odot}$ \citep{2009ApJ...692..538R}. It is a well studied BD in wavelengths ranging from X-rays to millimeters \citep{2001ApJ...561L.195M,2003ApJ...585..372L,2003ApJ...590L.111P,2003ApJ...593L..57K,2004A&A...426L..53A,2007A&A...468..391G}. Using the Submillimeter Common-User Bolometer Array (SCUBA) on James Clerk Maxwell Telescope (JCMT) and the Max-Planck Millimeter Bolometer (MAMBO) array on the IRAM 30 m telescope, \cite{2003ApJ...593L..57K} reported the presence of circumstellar cold dust around this young object. The spectral energy distribution (SED) of this circumstellar dust fits a flat disk model better than spherical dust distribution model and resembles to that of T Tauri disk with a mass estimation of (0.3-1.5)\textit{M$_{J}$} \citep{2003ApJ...590L.111P}. The mid-infrared observations done by using GEMINI/T-ReCS suggest the presence of 2$\mu$m silicon like grains is prominent in the disk \citep{2004A&A...426L..53A}. In addition, Atacama Large Millimeter/submillimeter Array (ALMA) observations at 0.89 mm and 3.2 mm show that large grains of at least $\sim$ 1 mm are also present in the outer disk region and the outer radius of disk is $>$80 AU  \citep{2014ApJ...791...20R}.
\\\\
\cite{2001ApJ...561L.195M} detected a strong H$\alpha$ emission with an equivalent width 340 \AA, and Br$\gamma$ emission from CT4.  Likewise, \cite{2007A&A...468..391G} detected quiescent X-rays with luminosity equal to 24.3 $\times$ 10$^{28}$ erg s$^{-1}$ and X-ray activity, log(L$_{X}$/L$_{bol}$) = -3 which is also the saturated X-ray emission level of early M dwarfs. Hence the two magnetic activity indicators H$\alpha$ emission and X-ray emission signify the presence of active chromoshphere and corona in this substellar object. \cite{2003ApJ...592..282J} classify CT4 as a non-accretor using the shape and width of H$\alpha$ emission profile but \cite{2009ApJ...697..373R} mention that it may have magnetospheric accretion. Different authors have reported different values of visual extinction parameter $A_{V}$ for CT4. \cite{2001ApJ...561L.195M} reported its value equal to 3.0 mag in their discovery paper, using the \textit{I-J} colors of field M dwarfs and the interstellar extinction law of \cite{1985ApJ...288..618R}. Likewise, \cite{2010A&A...515A..91M,2017AJ....153...46L,2012A&A...539A.151A,2013ApJ...771..129A} report its value equal to 2.6 mag, 5.0 mag, 5.4 mag and 5.67$\pm$0.89 mag respectively. \cite{2010A&A...515A..91M} do not specify clearly how they estimated $A_{V}$ particularly for CT4. \cite{2012A&A...539A.151A} estimated $A_{V}$ from the \textit{ J-H} vs. \textit{H-K$_{s}$} diagram. \cite{2017AJ....153...46L} used SpeX spectrum of CT4 and estimated the value of $A_{V}$ by comparing the spectral slopes at 1$\mu$m, of various young M dwarfs. Likewise, \cite{2013ApJ...771..129A} estimated $A_{V}$ by fitting the stellar photosphere models to the optical and near-infrared spectral energy distribution (SED) of the object. In addition to all the above values of $A_{V}$, \cite{2018arXiv180401533Z}  estimated it to be 6.37$\pm$0.85 mag. Their estimation is based on the intrinsic optical$\mbox{--}$near-infrared color as a function of spectral type.  They used spectral type of M7.2$\pm$0.9 for CT4, which was estimated by using their own reddening-free spectral classification system. As this value of $A_{V}$ is based on more precise photometry measured by Pan-STARRS1 3$\pi$ survey, we use it for estimation of flare energies in this paper\footnote{\cite{2018arXiv180401533Z} also estimated the value of $A_{V}$ based on color-color diagrams using H$_{2}$O indices but suggest that the value of $A_{V}$ based on intrinsic optical$\mbox{--}$near-infrared color sequences is more precise whenever the reddening-free spectral type of the object is well-defined, and $\approx$M5-L2.}. The photometric and physical properties of CT4 are summarized in Table \ref{table:target characteristics}.
\begin{table*}
 	\caption{Properties of CFHT-BD-Tau 4}
     \centering
     \begin{tabular}{cccc}
     \hline
     \hline
       Parameter & Value & Unit & Ref.\\
       
       \hline
        PHOTOMETRIC PARAMETERS\\
       \hline
       \hline
       Sp. Type & M7  & & 1\\
       \textit{V }& 21.556$\pm$0.008 & mag & 2 \\
       \textit{J }& 12.17$\pm$0.02 & mag & 3 \\
       	\textit{H} & 11.01$\pm$0.02 & mag & 3 \\
       	\textit{K$_{s}$} & 10.33$\pm$0.02 & mag & 3 \\
       	\textit{r} & 20.20$\pm$0.04 & mag & 4\\
       	\textit{i} & 17.54$\pm$0.02 & mag & 4\\
       	\textit{z} & 15.79$\pm$0.01 & mag & 4\\
       	\textit{G} & 17.780$\pm$0.006 & mag & 5 \\
       	\hline
       	\hline
       	PHYSICAL PARAMETERS\\
       	\hline
       	\hline
       	\textit{T$_{eff}$} & 2900 & K & 6\\
       	\textit{M} & 0.064\footnote{\cite{2014ApJ...791...20R} mention the mass of CT4 to be 0.095$M_{\odot}$, based on the parameters derived by \cite{2013ApJ...771..129A}.}  & $M_{\odot}$ & 6 \\
       	\textit{R }& 0.65 & $R_{\odot}$ & 6 \\
       	log \textit{L/L$_{\odot}$} & -1.57 & & 6\\
       	$v \sin i$ & 6$^{+2}_{-4}$ & km s$^{-1}$ & 6\\
       	Age & $\sim$1 & Myr & 7\\
       	parallax & 6.80$\pm$0.24 & mas & 5 \\
       $\alpha$ & 069.94787355129$^{b}$ ($\pm$0.2 mas) & deg & 5\\
        $\delta$ & +26.02787631047$^{b}$ ($\pm$0.1 mas) & deg & 5 \\
       	\hline
      	\end{tabular}
      	\\
      	$^{b}$epoch J2015.5, ICRS \\
      	{\textbf{References}}: \\
      	(1) \cite{2001ApJ...561L.195M}; (2) \cite{2006ApJ...649..306K}; \\(3) \cite{2006AJ....131.1163S};  (4) \cite{2016arXiv161205560C}; \\
      (5) \cite{2018arXiv180409365G};  (6) \cite{2009ApJ...697..373R}; \\ (7) \cite{2017AJ....153...46L} 
\end{table*}
 \label{table:target characteristics}
\section{Data reduction and Computations} \label{sec: data reduction and computations}
\subsection{\textit{Kepler K2} photometry}
 CFHT-BD-Tau 4 was observed as EPIC 248029954 by \textit{Kepler K2} mission during Campaign 13 (08 March, 2017  - 27 May, 2017) in long cadence ($\sim$30 minute) mode \citep{2010ApJ...713L.120J}. The total observation time is 80.52 day, and there are 3651 good (Q = 0) data points. Average flux measured during 29.4 minute interval is obtained for each data point. We used astropy affiliated photutils package to measure the photometry of flare from available Target Pixel File (TPF), by following a similar process described in \cite{2017ApJ...845...33G,2017ApJ...838...22G}. To get the best estimate of the position of target in each pixel image, we first estimated the median of centroids of all images. We then used the information recorded as POS\_CORR1 and POS\_CORR2 in the FITS file headers for each observation, to correct any off-positioning of the target centroid due to spacecraft motion. Based on experience in our previous works \cite{2017ApJ...838...22G,2017ApJ...845...33G}, this method is preferable for UCDs.\\\\
 We used \cite{2015ApJ...806...30L} relation: $\tilde{K_{p}}$ $\equiv$ 25.3 - 2.5log(count rate)  to estimate the \textit{Kepler} brightness magnitude $\tilde{K_{p}}$ which describes brightness level of the faint stars better than magnitude $K_{p}$ provided in \textit{Kepler} Input Catalog (KIC) \citep{2017ApJ...838...22G,2017ApJ...845...33G}. Here flux is the count rate measured through a 3-pixel radius aperture. In case of CFHT-BD-Tau 4, the median counts through 3-pixel radius aperture is less than through 2-pixel radius aperture due to the negative counts in the outer pixels surrounding the target pixel.  The median count rate through 2-pixel radius aperture is 478 count s$^{-1}$  which corresponds to photoshperic continuum level. This gives $\tilde{K_{p}}$ = 18.6. There is no significant difference in $\tilde{K_{p}}$ calculated using either 2-pixel or 3-pixel radius aperture in case of UCDs. We used 2-pixel radius aperture to measure the photometry of flares discussed in this paper. The \textit{K2} light curve of CT4 is shown in Figure \ref{fig:lightcurve} in which we can see periodic nature of the curve. Using Lomb-Scargle periodogram, we find that the period of this periodic feature is 2.98 days, and is comparable to rotation period of 2.93$\pm^{0.9}_{2.4}$ day reported by \cite{2018arXiv180407380S} for CT4. We compared the light curve  obtained by using aperture photometry with that obtained by using psf photometry and EVEREST detrending code \citep{2016AJ....152..100L}. We find no significant differences between the light curves.
 \begin{figure*} 
  \includegraphics[scale=0.45,angle=0]{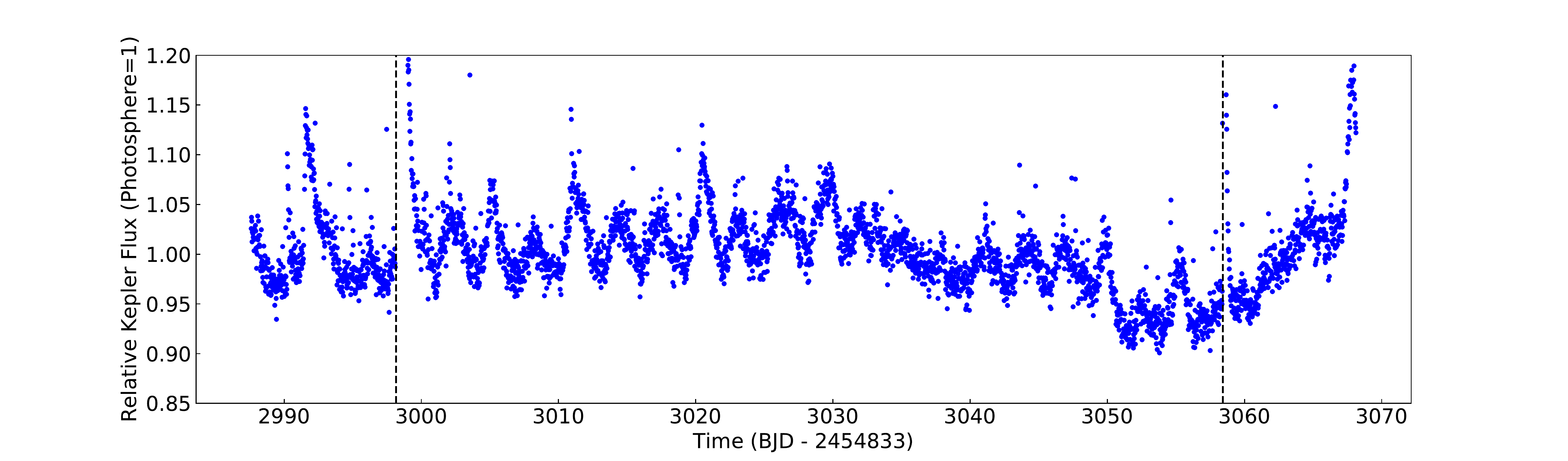} 
  \caption{The \textit{K2} light curve of CT4. The two superflares are not shown in full scale to focus the periodic nature of the curve. Using Lomb-Scargle periodogram, the dominant period of the light curve is 2.98 day. The two dashed vertical lines mark the peak flux times of the superflares.}
  \end{figure*}
  \label{fig:lightcurve}  
 \\
\subsection{Computation of Flare Energies}
To compute energy of a flare, we first estimated its equivalent duration (ED). It is the equivalent time during which the (sub)stellar object (in its quiescent state) would have emited the same amount of energy as the flare actually emitted \citep{1972Ap&SS..19...75G}. ED has units of time, and is area under the flare light curve. It depends on the filter used but is independent of the distance to the flaring object. However, it is affected by reddening. The quiescent bolometric luminosity of flare is calculated by approximating the flare to be a 10,000 K blackbody which produces the same count rate through the \textit{Kepler} response curve as CT4. We estimated the photospheric spectrum of CT4 by using an active M7 template spectrum \citep{2007AJ....133..531B} normalized to match the Pan-STARRS\textit{ i}-band photometry \citep{2012ApJ...750...99T,2016arXiv161205560C,2016arXiv161205242M}. As \textit{Kepler} flux of CT4 is affected by extinction, we first reddened the blackbody spectrum using astropy application specutil.extinction for $A_{V}$ = 6.37, and $R_{V}$ = 3.1. Here, the value of $R_{V}$(=$A_{V}/E(B-V))$ is the typical value used for Milky Way Galaxy. We used the mean extinction law as mentioned in \cite{1989ApJ...345..245C}. Then, we computed the photospheric flux of the reddened 10,000 K blackbody, which is normalized to have the same count rate through the \textit{Kepler} filter as the photosphere of CT4. The bolometric (UV/optical/IR) energy of CT4 flare having equivalent duration of 1 s is 4.9 $\times$ 10$^{30}$ erg and 5.4 $\times$ 10$^{32}$ erg corresponding to $A_{V}$ = 0.0 and 6.37 respectively. These energies were used to estimate bolometric flare energy by multiplying with ED of the flare. We adopted a distance of 147.1 pc to calculate the bolometric luminosity. More details of flare energy calibration can be found in \cite{2017ApJ...838...22G,2017ApJ...845...33G}. Figure \ref{fig:flare energy calibration} shows optical spectral energy distribution (SED) of CT4. For comparison, we also plot SEDs of 10,000 K and 6,500 K blackbody flares that produce the same counts as CT4 through \textit{Kepler} filter. The upper plot shows the reddened SEDs, and the lower plot shows the dereddened SEDs for two values of $A_{V}$: 6.37 and 3.0. 
\begin{figure} 
 \includegraphics[scale=0.50,angle=0]{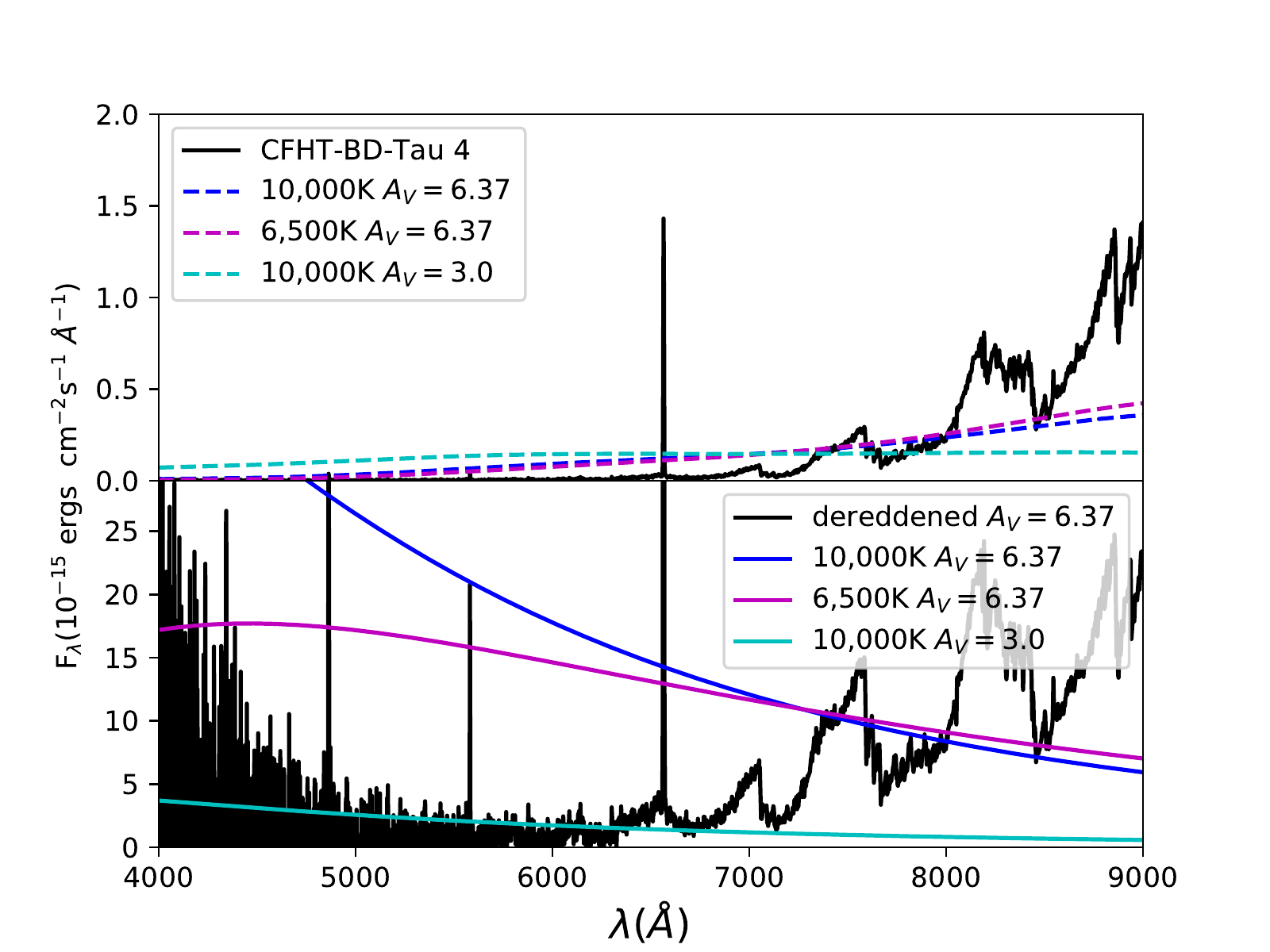} 
 \caption{Optical spectral energy distribution (SED) of CT4 (black solid curve). The three dashed lines represent the SEDs of hypothetical blackbody flares of temperature 10,000 K, 6,500 K and 10,000 K which produce the same counts through \textit{Kepler} filter, corresponding to \textit{A$_{V}$} = 6.37, 6.37 and 3.0 respectively. The upper plot is the reddened version of SEDs, and the lower plot is the dereddened version of SEDs.}
 \end{figure}    
\label{fig:flare energy calibration}
\subsection{Flare photometry}
\begin{figure*} 
\centering
 \includegraphics[scale=0.80,angle=0]{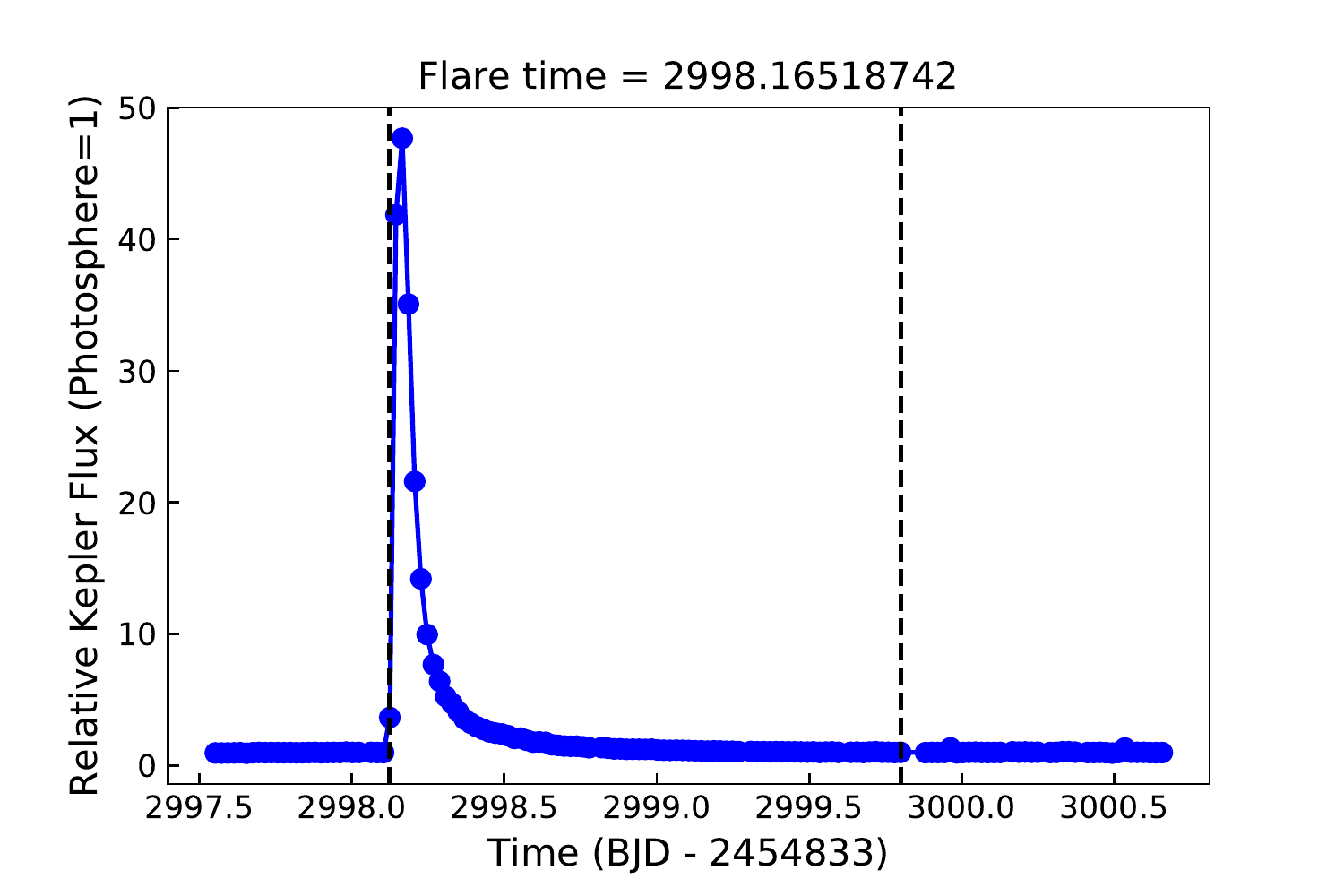} 
 \caption{The strongest superflare observed on CFHT-BD-Tau 4. The blue dots represent the observed data. The dashed vertical lines represent the start and end times of flare.}
 \end{figure*}    
\label{fig:big_flare} 
\begin{figure} 
 \includegraphics[scale=0.60,angle=0]{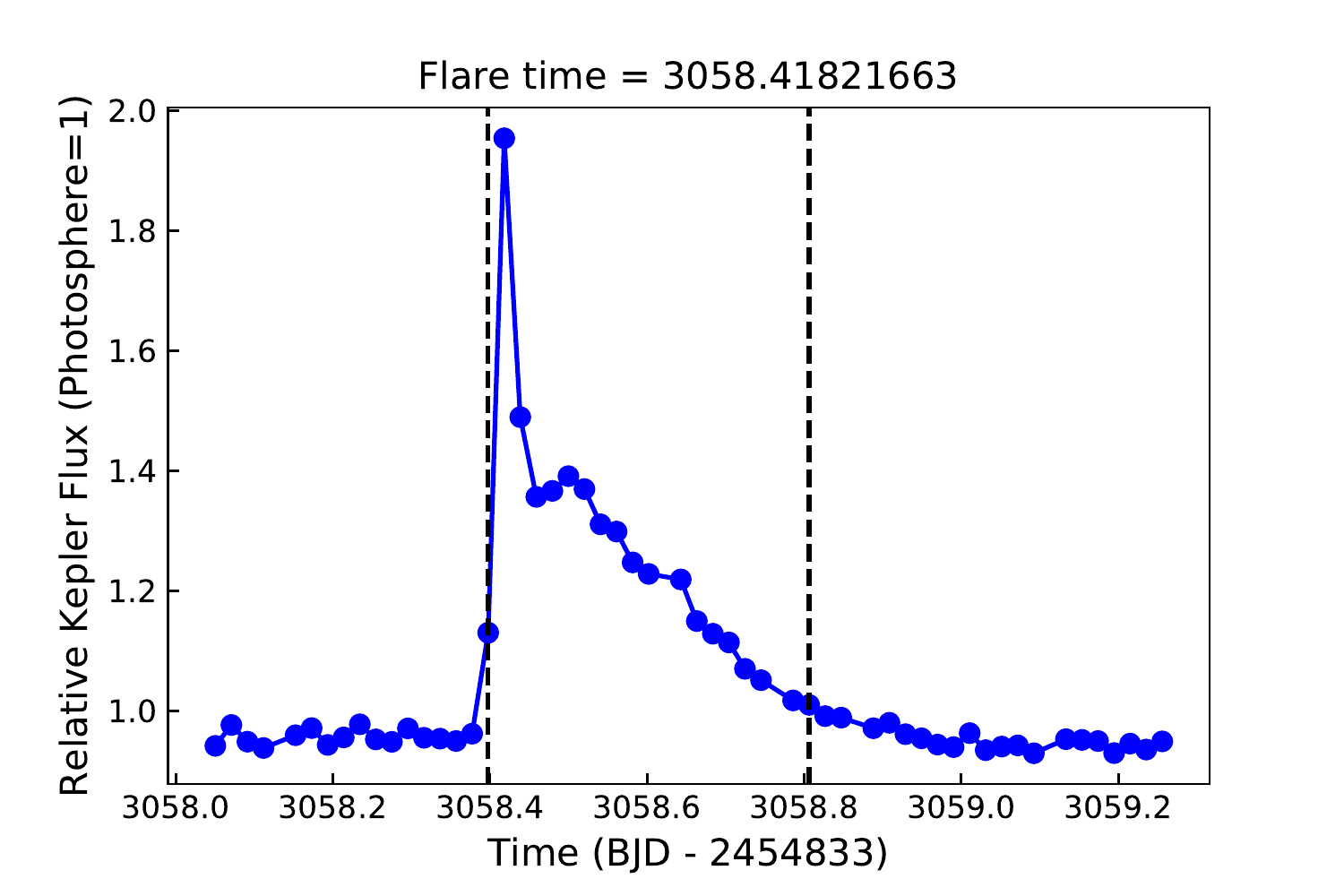} 
 \caption{Next superflare observed on CFHT-BD-Tau 4. The blue dots represent the observed data. The dashed vertical lines represent the start and end times of flare.}
 \end{figure}    
\label{fig:flare2} 
The strongest superflare detected on CT4  is shown in Figure~\ref{fig:big_flare}. At first, the flux increased to 1753 count s$^{-1}$ and 20006 count s$^{-1}$ at \textit{Kepler} mission day 2998.1243 and 2998.1447 respectively. It increased to peak value of 22788 count s$^{-1}$($\tilde{K_{p}}$ = 14.4, $\Delta \tilde{K_{p}}$ = -4.20) on \textit{Kelper} mission day 2998.1651, during which the target brightened by $\sim$ 48 times the quiescent photospheric level. Then, the flare continued to decay for over the next several hours as seen in Figure \ref{fig:big_flare}. This flare has an equivalent duration of $\sim$107 hour, and an estimated total bolometric (UV/optical/IR) energy equal to 2.1 $\times$ 10$^{38}$ erg, for $A_{V}$ = 6.37. The total flare duration is 1.7 day. The rise time of the flare is longer than that of superflares observed on some ultracool dwarfs. Further discussion is given in Section \ref{sec:discussion}. \\
\\
We also detected another superflare on CT4, as shown in Figure \ref{fig:flare2}. It is a complex flare with two peaks. The first peak is at \textit{Kepler} mission day 3058.4182, and the second peak is at \textit{Kepler} mission day 3058.4999. The flux counts at those times are 934 count s$^{-1}$ and 665 count s$^{-1}$ respectively. This flare has an ED of $\sim$2.4 hour, and flare duration of 0.41 day. The estimated total bolometric (UV/optical/IR) energy of this flare is 4.7 $\times$ 10$^{36}$ erg, for $A_{V}$ = 6.37. The properties of both flares are summarized in Table \ref{table:flare energies}. For comparison, we also list the flare energies for zero extinction i.e. $A_{V}$ = 0.0.
\begin{table*}
 	\caption{Flare properties}
     \centering
     \begin{tabular*}{0.65\textwidth}{cccccc}
     \hline
     \hline
        & ED (hour) & $A_{V}$ = 0.0 & $A_{V}$ = 6.37  & flare duration (day)\\
       \hline
       Flare 1 & 107 & 1.9 $\times$ 10$^{36}$ erg  &  2.1 $\times$ 10$^{38}$ erg  & 1.7 \\
       Flare 2 & 2.4 & 4.2 $\times$ 10$^{34}$ erg   & 4.7 $\times$ 10$^{36}$ erg & 0.41 \\
       	\hline
             	\end{tabular*}
\end{table*}
\label{table:flare energies}
\\
\section{Discussion} \label{sec:discussion}
We observed two superflares on CT4 using \textit{Kepler K2} Campaign 13 long cadence data. We estimated total bolometric (UV/optical/IR) energies of those flares to be 2.1 $\times$ 10$^{38}$ erg and 4.7 $\times$ 10$^{36}$ erg, for $A_{V}$ = 6.37. The stronger of the two superflares has ED  of $\sim$107 hour and the weaker has ED of $\sim$2.4 hour. The energies of these flares are larger than the strongest flares reported on other young brown dwarfs 2M0335+2342 and CFHT-PL-17 in \cite{2017ApJ...838...22G}. While the energy partition may not be same even for the flares that occur on same star \citep{2016ApJ...832..174O}, we may get an approximate estimation of soft X-ray (0.01-10 keV) energy (\textit{E$_{X}$}) radiated during the stronger superflare by using the conversion factors: \textit{E$_{X}$/E$_{bol}$} = 0.3, and \textit{E$_{Kp}$/E$_{bol}$} = 0.16 (assuming 9,000 K blackbody temperature) for active stars, listed in Table 2 of \cite{2015ApJ...809...79O}. Here, \textit{E$_{bol}$} is the total bolometric flare radiated energy, and is related to the coronal radiated flare energy \textit{E$_{cor}$} and the optical flare energy \textit{E$_{opt}$}, by \textit{E$_{bol}$} = \textit{E$_{cor}$ + E$_{opt}$}. Likewise, \textit{E$_{Kp}$} is the \textit{Kepler} flare energy. Our estimation of \textit{E$_{Kp}$} for the stronger superflare is 6.8 $\times$ 10$^{37}$ erg (assuming 10,000 K blackbody temperature). Using this value of \textit{E$_{Kp}$} and the conversion factors mentioned above, we get \textit{E$_{X}$} $\approx$ 1.3 $\times$ 10$^{38}$ erg. The occurence of those two superflares during the observed time of 78.31 days implies that CT4 is highly active which is also supported by its high H$\alpha$ and X-ray emission and it is very young. We observed only two flares on CT4 because it was observed in long cadence mode.  If we had short cadence ($\sim$1 min, \citealt{2010ApJ...713L.160G}) data for CT4, it would be possible to detect more weaker flares. Using the results of flare frequency distribution of 24 Myr brown dwarf 2M0335+2342 over the range 10$^{31}$ to 10$^{33}$ erg as reported by \cite{2017ApJ...845...33G}, the expected rate of flare with energy equal to that of stronger superflare observed on CT4 is 0.004 per year, much less than the observed results. It is interesting to note that the energies of superflares observed on CT4 are comparable to energy of the most powerful flare reported by \cite{1990IAUS..137..193G}. He studied the energies of 13 flares on seven T Tauri stars which were observed by using Str$\ddot{o}$mgren filters and reported an upper limit of flare energy $\sim$7.0 $\times$ 10$^{36}$ erg in optical spectral region. His results are based on ground based observations over several nights with as few as 3 observations per night. \\ \\
%
%
%
%
We emphasize that the bolometric flare energies reported in this paper have huge uncertainties for several reasons. The flare energies strongly depend on the value of $A_{V}$, as can be seen in Table \ref{table:flare energies}. The flares contribute more flux in the shorter wavelengths in \textit{Kepler} band pass, and the shorter wavelengths are more affected by extinction in interstellar medium. Consequently, fewer photons are recorded giving rise to shorter ED and less flare energy. Furthermore, we have not taken into account the contribution of atomic emission lines to total energy budget. The atomic emission lines and blackbody components contribute in different proportions during the impulsive and gradual phase of the flares \citep{1991ApJ...378..725H}. 
\\ \\
\cite{1990IAUS..137..193G} pointed out that the light curves of strong flares on T Tauri stars are characterized by a slow rise and then a slow decline in brightness. This is different than that is observed in flares of ordinary flare stars, most of which are characterized by a rapid rise followed by a gradual decline. While it would be more clear if we had short cadence data of CT4, the long cadence data somehow suggests similar characteristics as pointed by \cite{1990IAUS..137..193G}. It took one hour for the strongest flare to reach the peak flux level. The full width half maximum (FWHM) time scale of the flare is between 1.5 and 2.0 hours. This time scale is very different than that observed on typical superflares on older, diskless targets. The FWHM time scale of strongest flares on some UCDs is of the order of few minutes. For example, the two superflares observed on the L0 dwarf 2MASS J12321827-0951502 and the M7 dwarf 2MASS J08352366+1029318 had very short FWHM time scales of order $\sim$2 minutes \citep{2018arXiv180307708P}. We should note that 2MASS J12321827-0951502 and 2MASS J08352366+1029318 were observed in short cadence ($\sim$1 min, \citealt{2010ApJ...713L.160G}) mode. It is not clear whether the presence of disk changes the nature of flare light curves of young objects like CT4. \\ \\
Assuming a similarity between the solar flares and the flares on BDs, we may get a rough estimation of maximum magnetic field strength \textit{B$^{max}_{z}$} associated with the stronger superflare observed on CT4 using the scaling relation in \cite{2013A&A...549A..66A}:
\begin{equation}
E = 0.5 \times 10^{32} \Big(\frac{B^{max}_{z}}{1000 G}\Big)^{2}\Big(\frac{L^{bipole}} {50 Mm}\Big)^{3}erg.
\end{equation}
where \textit{E} is the bolometric flare energy, and \textit{L$^{bipole}$} is linear separation between bipoles. If we take \textit{L$^{bipole}$} to be equal to $\pi$$R$ as the maximum distance between a pair of magnetic poles on the surface of CT4,  a strong magnetic field of 13.5 kG is required to produce the stronger superflare with energy 2.1 $\times$ 10$^{38}$ erg. In general, it is possible for fully convective M dwarfs to have strong average magnetic fields with the highest observed value to be 7.0 kG in the case of WX Ursae Majoris (Gliese 412 B) \citep{2017NatAs...1E.184S}. Likewise, a $\sim$22 Myr old M8.5 brown dwarf LSR J1835+3259 is reported to have a strong magnetic field of strength 5 kG \citep{2017arXiv170902861B}. However, the above value of magnetic field strength estimated in case of CT4 using solar flare model is higher compared to results of \cite{2009ApJ...697..373R} who reported weak magnetic fields (few hundred gauss) with an upper limit $\sim$1 kG on four accreting brown dwarfs including CT4.  Using magnetospheric accretion model to young accreting brown dwarfs in combination with observed data, \cite{2006ApJ...638.1056S} and \cite{2007ApJ...671..842S} also predicted weak magnetic fields of  strength $\sim$kilogauss on the surface of those young objects. The longer FWHM  time scale of the superflare observed on CT4 might be possible due to the reason that it occured as a result of reconnection between the magnetic loops on CT4 and its disk. In such case, even the weak magnetic fields might be capable of producing superflares with huge energies because of the larger volumes associated with longer magnetic loops.\\
\\
It's possible that an outbrust from magnetospheric accretion could mimic a flare. Using magnetospheric model to a star-disk system, the rate of energy released \textit{$L_{acc}$} due to accretion of gas onto the star can be calculated by using 
\begin{equation}
L_{acc} =  \frac{GM_{*}\dot{M}}{R_{*}}\Bigg(1 - \frac{R_{*}}{R_{in}}\Bigg)
\end{equation}
where $M_{*}$ and $R_{*}$ are mass and radius of the star; $\dot{M}$ is the accretion rate, and $R_{in}$ is the inner radius of accretion disk \citep{1998ApJ...492..323G}. We do not have accurate estimation of inner disk radius and accretion rate of CT4. \cite{2003ApJ...585..372L} suggest $R_{in}$ $\sim$ (2-3)$R_{*}$ for young brown dwarfs. Their results are based on 38 cool objects in IC348 and Taurus which have spectral types M6-M9 and ages $\leq$5 Myr.  Using $R_{in}$ = 2.5$R_{*}$, an accretion rate log$\dot{M}$ = -6.70 $M_{\odot}$yr$^{-1}$ is required for the stronger superflare to emit the energy of 2.1 $\times$ 10$^{38}$ erg in the observed flare duration. This accretion rate is higher than those reported for young accreting BDs GY11 and 2MASS J053825.4-024241. \cite{2011A&A...526L...6R} reported an accretion rate of log$\dot{M}$ $\sim$ -9.86$\pm$0.45 $M_{\odot}$yr$^{-1}$ for 2MASS J053825.4-024241, with comparable mass, age and spectral type as CT4. Likewise, \cite{2010A&A...522A..47C} reported log$\dot{M}$ $\sim$-9.02 $M_{\odot}$yr$^{-1}$ for another deuterium burning brown dwarf GY 11. The higher value of requisite rate of accretion suggests that it is unlikely for the stronger superflare to occur due to magnetospheric accretion process.   \\ \\
The TRAPPIST-1 planetary system \citep{2016Natur.533..221G,2017Natur.542..456G,2017NatAs...1E.129L} demonstrates the existence of planets around low-mass stars and BDs, increasing the importance of the study of planet formation around low-mass stars and BDs. The superflares observed on CT4 will be very helpful to understand how such flares impact the dynamical and chemical evolution of disk around it. Such flares result to enhanced UV and X-ray emission which can create innerhole in the disk through photoevaporation \citep{2011MNRAS.412...13O}. The high energy X-rays also increase the ionization of the disk, which might trigger magnetorotational instability (MRI) that is supposed to drive magneto-hydrodynamical turbulence in the protoplanetary disk. This has several consequences on planet formation and depends on the energy of X-rays \citep{2010PNAS..107.7153F,2015ApJ...799..204C}. In addition, the study of superflares on young objects like CT4 might be helpful in explaining the mystries regarding the formation of the chondrules and the calcium-aluminium-rich inclusions (CAIs), which need transient heat sources to melt the precursor material. It is impossible to explain the formation of these materials in the context of thermodynamic equilibrium between the PMS stars and the disks around them. Some possible proposed transient heat sources are nebular lightning, protoplanetary induced shocks, activity associated with the young star having disk, and nearby Gamma Ray Burst (GRB) \citep{1999A&A...351..759M,2000Icar..143...87D,2001grba.conf..294D,2002ApJ...572..335F}. If strong flares acted as heating sources for formation of chondrules/CAIs, it is unclear how they were transported to the Asteroid belt \citep{2005ESASP.560..175F}.

\acknowledgements{Acknowledgements}\\
We are thankful to Zhoujian Zhang for providing us the results of his paper.
This paper makes use of data collected by the \textit{Kepler} mission. Funding for the \textit{Kepler} mission is provided by the NASA Science Mission derectorate. The material is based upon work supported by NASA under award Nos. NNX15AV64G, NNX16AE55G and NNX16AJ22G. P.K.G.W. and E.B. acknowledge support for this work from the National Science Foundation through Grant AST-1614770. Some/all of the data presented in this paper were obtained from the Mikulski Archive for Space Telescopes (MAST). STScI is operated by the Association of Universities for Research in Astronomy, Inc., under NASA contract NAS5-26555. Support for MAST for non-HST data is provided by the NASA Office of Space Science via grant NNX09AF08G and by other grants and contracts. This paper includes data collected by the Kepler mission. Funding for the Kepler mission is provided by the NASA Science Mission directorate.  This publication also
 makes use of data from the Sloan Digital Sky Survey. Funding
 for the Sloan Digital Sky Survey IV has been provided by the
Alfred P. Sloan Foundation, the U.S. Department of Energy
 Office of Science, and the Participating Institutions. SDSS-IV
 acknowledges support and resources from the Center for HighPerformance
 Computing at the University of Utah. The SDSS
 web site is www.sdss.org. SDSS-IV is managed by the
 Astrophysical Research Consortium for the Participating
 Institutions of the SDSS Collaboration, which can be found
 at http://www.sdss.org/collaboration/affiliations/. This work has made use of data from the European Space Agency (ESA) mission
 {\it Gaia} (\url{https://www.cosmos.esa.int/gaia}), processed by the {\it Gaia}
 Data Processing and Analysis Consortium (DPAC,
 \url{https://www.cosmos.esa.int/web/gaia/dpac/consortium}). Funding for the DPAC
 has been provided by national institutions, in particular the institutions
 participating in the {\it Gaia} Multilateral Agreement. This work made use of the \url{http://gaia-kepler.fun}  crossmatch database created by Megan Bedell.
 
\software: Astropy \citep{2013A&A...558A..33A}, Matplotlib \citep{Hunter:2007}, Numpy (Oliphant 2006)
\facility: 
\textit{Kepler}
\bibliographystyle{aasjournal}
\bibliography{/Users/rishipaudel/GoogleDrive/Research/astrobib}

\begin{thebibliography}{}
\expandafter\ifx\csname natexlab\endcsname\relax\def\natexlab#1{#1}\fi
\providecommand{\url}[1]{\href{#1}{#1}}

\bibitem[{{Alves de Oliveira} {et~al.}(2012){Alves de Oliveira}, {Moraux},
  {Bouvier}, \& {Bouy}}]{2012A&A...539A.151A}
{Alves de Oliveira}, C., {Moraux}, E., {Bouvier}, J., \& {Bouy}, H. 2012, \aap,
  539, A151

\bibitem[{{Andrews} {et~al.}(2013){Andrews}, {Rosenfeld}, {Kraus}, \&
  {Wilner}}]{2013ApJ...771..129A}
{Andrews}, S.~M., {Rosenfeld}, K.~A., {Kraus}, A.~L., \& {Wilner}, D.~J. 2013,
  \apj, 771, 129

\bibitem[{{Apai} {et~al.}(2004){Apai}, {Pascucci}, {Sterzik}, {van der Bliek},
  {Bouwman}, {Dullemond}, \& {Henning}}]{2004A&A...426L..53A}
{Apai}, D., {Pascucci}, I., {Sterzik}, M.~F., {et~al.} 2004, \aap, 426, L53

\bibitem[{{Astropy Collaboration} {et~al.}(2013){Astropy Collaboration},
  {Robitaille}, {Tollerud}, {Greenfield}, {Droettboom}, {Bray}, {Aldcroft},
  {Davis}, {Ginsburg}, {Price-Whelan}, {Kerzendorf}, {Conley}, {Crighton},
  {Barbary}, {Muna}, {Ferguson}, {Grollier}, {Parikh}, {Nair}, {Unther},
  {Deil}, {Woillez}, {Conseil}, {Kramer}, {Turner}, {Singer}, {Fox}, {Weaver},
  {Zabalza}, {Edwards}, {Azalee Bostroem}, {Burke}, {Casey}, {Crawford},
  {Dencheva}, {Ely}, {Jenness}, {Labrie}, {Lim}, {Pierfederici}, {Pontzen},
  {Ptak}, {Refsdal}, {Servillat}, \& {Streicher}}]{2013A&A...558A..33A}
{Astropy Collaboration}, {Robitaille}, T.~P., {Tollerud}, E.~J., {et~al.} 2013,
  \aap, 558, A33

\bibitem[{{Aulanier} {et~al.}(2013){Aulanier}, {D{\'e}moulin}, {Schrijver},
  {Janvier}, {Pariat}, \& {Schmieder}}]{2013A&A...549A..66A}
{Aulanier}, G., {D{\'e}moulin}, P., {Schrijver}, C.~J., {et~al.} 2013, \aap,
  549, A66

\bibitem[{{Berdyugina} {et~al.}(2017){Berdyugina}, {Harrington}, {Kuzmychov},
  {Kuhn}, {Hallinan}, {Kowalski}, \& {Hawley}}]{2017arXiv170902861B}
{Berdyugina}, S.~V., {Harrington}, D.~M., {Kuzmychov}, O., {et~al.} 2017, ArXiv
  e-prints, arXiv:1709.02861

\bibitem[{{Bochanski} {et~al.}(2007){Bochanski}, {West}, {Hawley}, \&
  {Covey}}]{2007AJ....133..531B}
{Bochanski}, J.~J., {West}, A.~A., {Hawley}, S.~L., \& {Covey}, K.~R. 2007,
  \aj, 133, 531

\bibitem[{{Candelaresi} {et~al.}(2014){Candelaresi}, {Hillier}, {Maehara},
  {Brandenburg}, \& {Shibata}}]{2014ApJ...792...67C}
{Candelaresi}, S., {Hillier}, A., {Maehara}, H., {Brandenburg}, A., \&
  {Shibata}, K. 2014, \apj, 792, 67

\bibitem[{{Cardelli} {et~al.}(1989){Cardelli}, {Clayton}, \&
  {Mathis}}]{1989ApJ...345..245C}
{Cardelli}, J.~A., {Clayton}, G.~C., \& {Mathis}, J.~S. 1989, \apj, 345, 245

\bibitem[{{Chambers} {et~al.}(2016){Chambers}, {Magnier}, {Metcalfe},
  {Flewelling}, {Huber}, {Waters}, {Denneau}, {Draper}, {Farrow}, {Finkbeiner},
  {Holmberg}, {Koppenhoefer}, {Price}, {Saglia}, {Schlafly}, {Smartt},
  {Sweeney}, {Wainscoat}, {Burgett}, {Grav}, {Heasley}, {Hodapp}, {Jedicke},
  {Kaiser}, {Kudritzki}, {Luppino}, {Lupton}, {Monet}, {Morgan}, {Onaka},
  {Stubbs}, {Tonry}, {Banados}, {Bell}, {Bender}, {Bernard}, {Botticella},
  {Casertano}, {Chastel}, {Chen}, {Chen}, {Cole}, {Deacon}, {Frenk},
  {Fitzsimmons}, {Gezari}, {Goessl}, {Goggia}, {Goldman}, {Grebel}, {Hambly},
  {Hasinger}, {Heavens}, {Heckman}, {Henderson}, {Henning}, {Holman}, {Hopp},
  {Ip}, {Isani}, {Keyes}, {Koekemoer}, {Kotak}, {Long}, {Lucey}, {Liu},
  {Martin}, {McLean}, {Morganson}, {Murphy}, {Nieto-Santisteban}, {Norberg},
  {Peacock}, {Pier}, {Postman}, {Primak}, {Rae}, {Rest}, {Riess}, {Riffeser},
  {Rix}, {Roser}, {Schilbach}, {Schultz}, {Scolnic}, {Szalay}, {Seitz},
  {Shiao}, {Small}, {Smith}, {Soderblom}, {Taylor}, {Thakar}, {Thiel},
  {Thilker}, {Urata}, {Valenti}, {Walter}, {Watters}, {Werner}, {White},
  {Wood-Vasey}, \& {Wyse}}]{2016arXiv161205560C}
{Chambers}, K.~C., {Magnier}, E.~A., {Metcalfe}, N., {et~al.} 2016, ArXiv
  e-prints, arXiv:1612.05560

\bibitem[{{Christensen} {et~al.}(2009){Christensen}, {Holzwarth}, \&
  {Reiners}}]{2009Natur.457..167C}
{Christensen}, U.~R., {Holzwarth}, V., \& {Reiners}, A. 2009, \nat, 457, 167

\bibitem[{{Cleeves} {et~al.}(2015){Cleeves}, {Bergin}, {Qi}, {Adams}, \&
  {{\"O}berg}}]{2015ApJ...799..204C}
{Cleeves}, L.~I., {Bergin}, E.~A., {Qi}, C., {Adams}, F.~C., \& {{\"O}berg},
  K.~I. 2015, \apj, 799, 204

\bibitem[{{Comer{\'o}n} {et~al.}(2010){Comer{\'o}n}, {Testi}, \&
  {Natta}}]{2010A&A...522A..47C}
{Comer{\'o}n}, F., {Testi}, L., \& {Natta}, A. 2010, \aap, 522, A47

\bibitem[{{Desch} \& {Cuzzi}(2000)}]{2000Icar..143...87D}
{Desch}, S.~J., \& {Cuzzi}, J.~N. 2000, \icarus, 143, 87

\bibitem[{{Duggan} {et~al.}(2001){Duggan}, {McBreen}, {Hanlon}, {Metcalfe},
  {Kvick}, \& {Vaughan}}]{2001grba.conf..294D}
{Duggan}, P., {McBreen}, B., {Hanlon}, L., {et~al.} 2001, in Gamma-ray Bursts
  in the Afterglow Era, ed. E.~{Costa}, F.~{Frontera}, \& J.~{Hjorth}, 294

\bibitem[{{Feigelson}(2005)}]{2005ESASP.560..175F}
{Feigelson}, E.~D. 2005, in ESA Special Publication, Vol. 560, 13th Cambridge
  Workshop on Cool Stars, Stellar Systems and the Sun, ed. F.~{Favata},
  G.~A.~J. {Hussain}, \& B.~{Battrick}, 175

\bibitem[{{Feigelson}(2010)}]{2010PNAS..107.7153F}
{Feigelson}, E.~D. 2010, Proceedings of the National Academy of Science, 107,
  7153

\bibitem[{{Feigelson} {et~al.}(2002){Feigelson}, {Garmire}, \&
  {Pravdo}}]{2002ApJ...572..335F}
{Feigelson}, E.~D., {Garmire}, G.~P., \& {Pravdo}, S.~H. 2002, \apj, 572, 335

\bibitem[{{Gahm}(1990)}]{1990IAUS..137..193G}
{Gahm}, G.~F. 1990, in IAU Symposium, Vol. 137, Flare Stars in Star Clusters,
  Associations and the Solar Vicinity, ed. L.~V. {Mirzoian}, B.~R. {Pettersen},
  \& M.~K. {Tsvetkov}, 193--206

\bibitem[{{Gaia Collaboration} {et~al.}(2018){Gaia Collaboration}, {Brown},
  {Vallenari}, {Prusti}, {de Bruijne}, {Babusiaux}, \&
  {Bailer-Jones}}]{2018arXiv180409365G}
{Gaia Collaboration}, {Brown}, A.~G.~A., {Vallenari}, A., {et~al.} 2018, ArXiv
  e-prints, arXiv:1804.09365

\bibitem[{{Gaia Collaboration} {et~al.}(2016){Gaia Collaboration}, {Prusti},
  {de Bruijne}, {Brown}, {Vallenari}, {Babusiaux}, {Bailer-Jones}, {Bastian},
  {Biermann}, {Evans}, \& et~al.}]{2016A&A...595A...1G}
{Gaia Collaboration}, {Prusti}, T., {de Bruijne}, J.~H.~J., {et~al.} 2016,
  \aap, 595, A1

\bibitem[{{Gershberg}(1972)}]{1972Ap&SS..19...75G}
{Gershberg}, R.~E. 1972, \apss, 19, 75

\bibitem[{{Getman} {et~al.}(2008{\natexlab{a}}){Getman}, {Feigelson}, {Broos},
  {Micela}, \& {Garmire}}]{2008ApJ...688..418G}
{Getman}, K.~V., {Feigelson}, E.~D., {Broos}, P.~S., {Micela}, G., \&
  {Garmire}, G.~P. 2008{\natexlab{a}}, \apj, 688, 418

\bibitem[{{Getman} {et~al.}(2008{\natexlab{b}}){Getman}, {Feigelson}, {Micela},
  {Jardine}, {Gregory}, \& {Garmire}}]{2008ApJ...688..437G}
{Getman}, K.~V., {Feigelson}, E.~D., {Micela}, G., {et~al.} 2008{\natexlab{b}},
  \apj, 688, 437

\bibitem[{{Gilliland} {et~al.}(2010){Gilliland}, {Jenkins}, {Borucki},
  {Bryson}, {Caldwell}, {Clarke}, {Dotson}, {Haas}, {Hall}, {Klaus}, {Koch},
  {McCauliff}, {Quintana}, {Twicken}, \& {van Cleve}}]{2010ApJ...713L.160G}
{Gilliland}, R.~L., {Jenkins}, J.~M., {Borucki}, W.~J., {et~al.} 2010, \apjl,
  713, L160

\bibitem[{{Gillon} {et~al.}(2016){Gillon}, {Jehin}, {Lederer}, {Delrez}, {de
  Wit}, {Burdanov}, {Van Grootel}, {Burgasser}, {Triaud}, {Opitom}, {Demory},
  {Sahu}, {Bardalez Gagliuffi}, {Magain}, \& {Queloz}}]{2016Natur.533..221G}
{Gillon}, M., {Jehin}, E., {Lederer}, S.~M., {et~al.} 2016, \nat, 533, 221

\bibitem[{{Gillon} {et~al.}(2017){Gillon}, {Triaud}, {Demory}, {Jehin}, {Agol},
  {Deck}, {Lederer}, {de Wit}, {Burdanov}, {Ingalls}, {Bolmont}, {Leconte},
  {Raymond}, {Selsis}, {Turbet}, {Barkaoui}, {Burgasser}, {Burleigh}, {Carey},
  {Chaushev}, {Copperwheat}, {Delrez}, {Fernandes}, {Holdsworth}, {Kotze}, {Van
  Grootel}, {Almleaky}, {Benkhaldoun}, {Magain}, \&
  {Queloz}}]{2017Natur.542..456G}
{Gillon}, M., {Triaud}, A.~H.~M.~J., {Demory}, B.-O., {et~al.} 2017, \nat, 542,
  456

\bibitem[{{Gizis} {et~al.}(2017{\natexlab{a}}){Gizis}, {Paudel}, {Mullan},
  {Schmidt}, {Burgasser}, \& {Williams}}]{2017ApJ...845...33G}
{Gizis}, J.~E., {Paudel}, R.~R., {Mullan}, D., {et~al.} 2017{\natexlab{a}},
  \apj, 845, 33

\bibitem[{{Gizis} {et~al.}(2017{\natexlab{b}}){Gizis}, {Paudel}, {Schmidt},
  {Williams}, \& {Burgasser}}]{2017ApJ...838...22G}
{Gizis}, J.~E., {Paudel}, R.~R., {Schmidt}, S.~J., {Williams}, P.~K.~G., \&
  {Burgasser}, A.~J. 2017{\natexlab{b}}, \apj, 838, 22

\bibitem[{{Grosso} {et~al.}(2007){Grosso}, {Briggs}, {G{\"u}del}, {Guieu},
  {Franciosini}, {Palla}, {Dougados}, {Monin}, {M{\'e}nard}, {Bouvier},
  {Audard}, \& {Telleschi}}]{2007A&A...468..391G}
{Grosso}, N., {Briggs}, K.~R., {G{\"u}del}, M., {et~al.} 2007, \aap, 468, 391

\bibitem[{{G{\"u}del} {et~al.}(2007){G{\"u}del}, {Briggs}, {Arzner}, {Audard},
  {Bouvier}, {Feigelson}, {Franciosini}, {Glauser}, {Grosso}, {Micela},
  {Monin}, {Montmerle}, {Padgett}, {Palla}, {Pillitteri}, {Rebull}, {Scelsi},
  {Silva}, {Skinner}, {Stelzer}, \& {Telleschi}}]{2007A&A...468..353G}
{G{\"u}del}, M., {Briggs}, K.~R., {Arzner}, K., {et~al.} 2007, \aap, 468, 353

\bibitem[{{Gullbring} {et~al.}(1998){Gullbring}, {Hartmann}, {Brice{\~n}o}, \&
  {Calvet}}]{1998ApJ...492..323G}
{Gullbring}, E., {Hartmann}, L., {Brice{\~n}o}, C., \& {Calvet}, N. 1998, \apj,
  492, 323

\bibitem[{{Hawley} \& {Pettersen}(1991)}]{1991ApJ...378..725H}
{Hawley}, S.~L., \& {Pettersen}, B.~R. 1991, \apj, 378, 725

\bibitem[{{Howell} {et~al.}(2014){Howell}, {Sobeck}, {Haas}, {Still},
  {Barclay}, {Mullally}, {Troeltzsch}, {Aigrain}, {Bryson}, {Caldwell},
  {Chaplin}, {Cochran}, {Huber}, {Marcy}, {Miglio}, {Najita}, {Smith},
  {Twicken}, \& {Fortney}}]{2014PASP..126..398H}
{Howell}, S.~B., {Sobeck}, C., {Haas}, M., {et~al.} 2014, \pasp, 126, 398

\bibitem[{Hunter(2007)}]{Hunter:2007}
Hunter, J.~D. 2007, Computing In Science \& Engineering, 9, 90

\bibitem[{{Jayawardhana} {et~al.}(2003){Jayawardhana}, {Mohanty}, \&
  {Basri}}]{2003ApJ...592..282J}
{Jayawardhana}, R., {Mohanty}, S., \& {Basri}, G. 2003, \apj, 592, 282

\bibitem[{{Jenkins} {et~al.}(2010){Jenkins}, {Caldwell}, {Chandrasekaran},
  {Twicken}, {Bryson}, {Quintana}, {Clarke}, {Li}, {Allen}, {Tenenbaum}, {Wu},
  {Klaus}, {Van Cleve}, {Dotson}, {Haas}, {Gilliland}, {Koch}, \&
  {Borucki}}]{2010ApJ...713L.120J}
{Jenkins}, J.~M., {Caldwell}, D.~A., {Chandrasekaran}, H., {et~al.} 2010,
  \apjl, 713, L120

\bibitem[{{Klein} {et~al.}(2003){Klein}, {Apai}, {Pascucci}, {Henning}, \&
  {Waters}}]{2003ApJ...593L..57K}
{Klein}, R., {Apai}, D., {Pascucci}, I., {Henning}, T., \& {Waters},
  L.~B.~F.~M. 2003, \apjl, 593, L57

\bibitem[{{Kraus} {et~al.}(2006){Kraus}, {White}, \&
  {Hillenbrand}}]{2006ApJ...649..306K}
{Kraus}, A.~L., {White}, R.~J., \& {Hillenbrand}, L.~A. 2006, \apj, 649, 306

\bibitem[{{Liu} {et~al.}(2003){Liu}, {Najita}, \&
  {Tokunaga}}]{2003ApJ...585..372L}
{Liu}, M.~C., {Najita}, J., \& {Tokunaga}, A.~T. 2003, \apj, 585, 372

\bibitem[{{Luger} {et~al.}(2016){Luger}, {Agol}, {Kruse}, {Barnes}, {Becker},
  {Foreman-Mackey}, \& {Deming}}]{2016AJ....152..100L}
{Luger}, R., {Agol}, E., {Kruse}, E., {et~al.} 2016, \aj, 152, 100

\bibitem[{{Luger} {et~al.}(2017){Luger}, {Sestovic}, {Kruse}, {Grimm},
  {Demory}, {Agol}, {Bolmont}, {Fabrycky}, {Fernandes}, {Van Grootel},
  {Burgasser}, {Gillon}, {Ingalls}, {Jehin}, {Raymond}, {Selsis}, {Triaud},
  {Barclay}, {Barentsen}, {Howell}, {Delrez}, {de Wit}, {Foreman-Mackey},
  {Holdsworth}, {Leconte}, {Lederer}, {Turbet}, {Almleaky}, {Benkhaldoun},
  {Magain}, {Morris}, {Heng}, \& {Queloz}}]{2017NatAs...1E.129L}
{Luger}, R., {Sestovic}, M., {Kruse}, E., {et~al.} 2017, Nature Astronomy, 1,
  0129

\bibitem[{{Luhman} {et~al.}(2017){Luhman}, {Mamajek}, {Shukla}, \&
  {Loutrel}}]{2017AJ....153...46L}
{Luhman}, K.~L., {Mamajek}, E.~E., {Shukla}, S.~J., \& {Loutrel}, N.~P. 2017,
  \aj, 153, 46

\bibitem[{{Lund} {et~al.}(2015){Lund}, {Handberg}, {Davies}, {Chaplin}, \&
  {Jones}}]{2015ApJ...806...30L}
{Lund}, M.~N., {Handberg}, R., {Davies}, G.~R., {Chaplin}, W.~J., \& {Jones},
  C.~D. 2015, \apj, 806, 30

\bibitem[{{Maehara} {et~al.}(2012){Maehara}, {Shibayama}, {Notsu}, {Notsu},
  {Nagao}, {Kusaba}, {Honda}, {Nogami}, \& {Shibata}}]{2012Natur.485..478M}
{Maehara}, H., {Shibayama}, T., {Notsu}, S., {et~al.} 2012, \nat, 485, 478

\bibitem[{{Magnier} {et~al.}(2016){Magnier}, {Schlafly}, {Finkbeiner}, {Tonry},
  {Goldman}, {R{\"o}ser}, {Schilbach}, {Chambers}, {Flewelling}, {Huber},
  {Price}, {Sweeney}, {Waters}, {Denneau}, {Draper}, {Hodapp}, {Jedicke},
  {Kudritzki}, {Metcalfe}, {Stubbs}, \& {Wainscoast}}]{2016arXiv161205242M}
{Magnier}, E.~A., {Schlafly}, E.~F., {Finkbeiner}, D.~P., {et~al.} 2016, ArXiv
  e-prints, arXiv:1612.05242

\bibitem[{{Mart{\'{\i}}n} {et~al.}(2001){Mart{\'{\i}}n}, {Dougados}, {Magnier},
  {M{\'e}nard}, {Magazz{\`u}}, {Cuillandre}, \&
  {Delfosse}}]{2001ApJ...561L.195M}
{Mart{\'{\i}}n}, E.~L., {Dougados}, C., {Magnier}, E., {et~al.} 2001, \apjl,
  561, L195

\bibitem[{{McBreen} \& {Hanlon}(1999)}]{1999A&A...351..759M}
{McBreen}, B., \& {Hanlon}, L. 1999, \aap, 351, 759

\bibitem[{{Monin} {et~al.}(2010){Monin}, {Guieu}, {Pinte}, {Rebull},
  {Goldsmith}, {Fukagawa}, {M{\'e}nard}, {Padgett}, {Stappelfeld}, {McCabe},
  {Carey}, {Noriega-Crespo}, {Brooke}, {Huard}, {Terebey}, {Hillenbrand}, \&
  {Guedel}}]{2010A&A...515A..91M}
{Monin}, J.-L., {Guieu}, S., {Pinte}, C., {et~al.} 2010, \aap, 515, A91

\bibitem[{{Notsu} {et~al.}(2013){Notsu}, {Shibayama}, {Maehara}, {Notsu},
  {Nagao}, {Honda}, {Ishii}, {Nogami}, \& {Shibata}}]{2013ApJ...771..127N}
{Notsu}, Y., {Shibayama}, T., {Maehara}, H., {et~al.} 2013, \apj, 771, 127

\bibitem[{{Osten} \& {Wolk}(2015)}]{2015ApJ...809...79O}
{Osten}, R.~A., \& {Wolk}, S.~J. 2015, \apj, 809, 79

\bibitem[{{Osten} {et~al.}(2016){Osten}, {Kowalski}, {Drake}, {Krimm}, {Page},
  {Gazeas}, {Kennea}, {Oates}, {Page}, {de Miguel}, {Nov{\'a}k}, {Apeltauer},
  \& {Gehrels}}]{2016ApJ...832..174O}
{Osten}, R.~A., {Kowalski}, A., {Drake}, S.~A., {et~al.} 2016, \apj, 832, 174

\bibitem[{{Owen} {et~al.}(2011){Owen}, {Ercolano}, \&
  {Clarke}}]{2011MNRAS.412...13O}
{Owen}, J.~E., {Ercolano}, B., \& {Clarke}, C.~J. 2011, \mnras, 412, 13

\bibitem[{{Pascucci} {et~al.}(2003){Pascucci}, {Apai}, {Henning}, \&
  {Dullemond}}]{2003ApJ...590L.111P}
{Pascucci}, I., {Apai}, D., {Henning}, T., \& {Dullemond}, C.~P. 2003, \apjl,
  590, L111

\bibitem[{{Paudel} {et~al.}(2018){Paudel}, {Gizis}, {Mullan}, {Schmidt},
  {Burgasser}, {Williams}, \& {Berger}}]{2018arXiv180307708P}
{Paudel}, R.~R., {Gizis}, J.~E., {Mullan}, D.~J., {et~al.} 2018, ArXiv
  e-prints, arXiv:1803.07708

\bibitem[{{Preibisch} {et~al.}(2005){Preibisch}, {McCaughrean}, {Grosso},
  {Feigelson}, {Flaccomio}, {Getman}, {Hillenbrand}, {Meeus}, {Micela},
  {Sciortino}, \& {Stelzer}}]{2005ApJS..160..582P}
{Preibisch}, T., {McCaughrean}, M.~J., {Grosso}, N., {et~al.} 2005, \apjs, 160,
  582

\bibitem[{{Reiners} {et~al.}(2009{\natexlab{a}}){Reiners}, {Basri}, \&
  {Browning}}]{2009ApJ...692..538R}
{Reiners}, A., {Basri}, G., \& {Browning}, M. 2009{\natexlab{a}}, \apj, 692,
  538

\bibitem[{{Reiners} {et~al.}(2009{\natexlab{b}}){Reiners}, {Basri}, \&
  {Christensen}}]{2009ApJ...697..373R}
{Reiners}, A., {Basri}, G., \& {Christensen}, U.~R. 2009{\natexlab{b}}, \apj,
  697, 373

\bibitem[{{Reiners} \& {Christensen}(2010)}]{2010A&A...522A..13R}
{Reiners}, A., \& {Christensen}, U.~R. 2010, \aap, 522, A13

\bibitem[{{Ricci} {et~al.}(2014){Ricci}, {Testi}, {Natta}, {Scholz}, {de
  Gregorio-Monsalvo}, \& {Isella}}]{2014ApJ...791...20R}
{Ricci}, L., {Testi}, L., {Natta}, A., {et~al.} 2014, \apj, 791, 20

\bibitem[{{Rieke} \& {Lebofsky}(1985)}]{1985ApJ...288..618R}
{Rieke}, G.~H., \& {Lebofsky}, M.~J. 1985, \apj, 288, 618

\bibitem[{{Rigliaco} {et~al.}(2011){Rigliaco}, {Natta}, {Randich}, {Testi},
  {Covino}, {Herczeg}, \& {Alcal{\'a}}}]{2011A&A...526L...6R}
{Rigliaco}, E., {Natta}, A., {Randich}, S., {et~al.} 2011, \aap, 526, L6

\bibitem[{{Scholz} \& {Jayawardhana}(2006)}]{2006ApJ...638.1056S}
{Scholz}, A., \& {Jayawardhana}, R. 2006, \apj, 638, 1056

\bibitem[{{Scholz} {et~al.}(2018){Scholz}, {Moore}, {Jayawardhana}, {Aigrain},
  {Peterson}, \& {Stelzer}}]{2018arXiv180407380S}
{Scholz}, A., {Moore}, K., {Jayawardhana}, R., {et~al.} 2018, ArXiv e-prints,
  arXiv:1804.07380

\bibitem[{{Shibayama} {et~al.}(2013){Shibayama}, {Maehara}, {Notsu}, {Notsu},
  {Nagao}, {Honda}, {Ishii}, {Nogami}, \& {Shibata}}]{2013ApJS..209....5S}
{Shibayama}, T., {Maehara}, H., {Notsu}, S., {et~al.} 2013, \apjs, 209, 5

\bibitem[{{Shulyak} {et~al.}(2017){Shulyak}, {Reiners}, {Engeln}, {Malo},
  {Yadav}, {Morin}, \& {Kochukhov}}]{2017NatAs...1E.184S}
{Shulyak}, D., {Reiners}, A., {Engeln}, A., {et~al.} 2017, Nature Astronomy, 1,
  0184

\bibitem[{{Skrutskie} {et~al.}(2006){Skrutskie}, {Cutri}, {Stiening},
  {Weinberg}, {Schneider}, {Carpenter}, {Beichman}, {Capps}, {Chester},
  {Elias}, {Huchra}, {Liebert}, {Lonsdale}, {Monet}, {Price}, {Seitzer},
  {Jarrett}, {Kirkpatrick}, {Gizis}, {Howard}, {Evans}, {Fowler}, {Fullmer},
  {Hurt}, {Light}, {Kopan}, {Marsh}, {McCallon}, {Tam}, {Van Dyk}, \&
  {Wheelock}}]{2006AJ....131.1163S}
{Skrutskie}, M.~F., {Cutri}, R.~M., {Stiening}, R., {et~al.} 2006, \aj, 131,
  1163

\bibitem[{{Stelzer} {et~al.}(2000){Stelzer}, {Neuh{\"a}user}, \&
  {Hambaryan}}]{2000A&A...356..949S}
{Stelzer}, B., {Neuh{\"a}user}, R., \& {Hambaryan}, V. 2000, \aap, 356, 949

\bibitem[{{Stelzer} {et~al.}(2007){Stelzer}, {Scholz}, \&
  {Jayawardhana}}]{2007ApJ...671..842S}
{Stelzer}, B., {Scholz}, A., \& {Jayawardhana}, R. 2007, \apj, 671, 842

\bibitem[{{Tonry} {et~al.}(2012){Tonry}, {Stubbs}, {Lykke}, {Doherty},
  {Shivvers}, {Burgett}, {Chambers}, {Hodapp}, {Kaiser}, {Kudritzki},
  {Magnier}, {Morgan}, {Price}, \& {Wainscoat}}]{2012ApJ...750...99T}
{Tonry}, J.~L., {Stubbs}, C.~W., {Lykke}, K.~R., {et~al.} 2012, \apj, 750, 99

\bibitem[{{Wolk} {et~al.}(2005){Wolk}, {Harnden}, {Flaccomio}, {Micela},
  {Favata}, {Shang}, \& {Feigelson}}]{2005ApJS..160..423W}
{Wolk}, S.~J., {Harnden}, Jr., F.~R., {Flaccomio}, E., {et~al.} 2005, \apjs,
  160, 423

\bibitem[{{Zhang} {et~al.}(2018){Zhang}, {Liu}, {Best}, {Magnier}, {Aller},
  {Chambers}, {Draper}, {Flewelling}, {Hodapp}, {Kaiser}, {Kudritzki},
  {Metcalfe}, {Wainscoat}, \& {Waters}}]{2018arXiv180401533Z}
{Zhang}, Z., {Liu}, M.~C., {Best}, W.~M.~J., {et~al.} 2018, ArXiv e-prints,
  arXiv:1804.01533

\bibitem[{{Zhu} {et~al.}(2009){Zhu}, {Hartmann}, \&
  {Gammie}}]{2009ApJ...694.1045Z}
{Zhu}, Z., {Hartmann}, L., \& {Gammie}, C. 2009, \apj, 694, 1045

\end{thebibliography}
 \end{document}